# NON-THERMAL FUSION BURNING PROCESSES, RELEVANT COLLECTIVE MODES AND GAINED PERSPECTIVES

B. COPPI
Massachusetts Institute of Technology and CNR
Cambridge, USA and Rome, Italy
Email: coppi@mit.edu

B. BASU
Massachusetts Institute of Technology
Cambridge, USA

**Abstract**

New tridimensional plasma structures, that are oscillatory and classified as non-separable ballooning modes, can emerge in inhomogeneous plasmas and undergo resonant mode-particle interactions, e.g., with a minority population, that can lead them to modify their spatial profiles. Thus, unlike the case of previously known ballooning modes their amplitudes are not separable functions of time and space. The relevant resonance conditions are intrinsically different from those of the well-known Landau conditions for (ordinary) plasma waves: they involve the mode geometry and affect different regions of the distribution in momentum space at different positions in configuration space. A process for a transfer of energy among different particle populations is envisioned.

1. INTRODUCTION

The same factors that make the physics of magnetically confined fusion burning plasmas difficult to predict provide also the opportunity to identify novel processes extending the range of meaningful fusion burn conditions which can be achieved beyond those predicted on the basis of (conventional) thermonuclear fusion theory.

In fact, weakly collisional and well confined plasmas have been found to be strongly influenced by the presence of collective modes and self-organization processes [1]. In fusion burning regimes, where the plasma energy is supplied by the charged reaction products, self-organization is expected to be more important than in present and past experiments where plasma heating is provided by an external and controlled source. Then new modes or new forms of previously known modes can be expected to emerge. In particular, given the relevance of novel [2] resonant mode-particle interactions [3], it is reasonable to expect that the distributions of the reacting nuclei in momentum space will not remain strictly Maxwellian and that the resulting reaction rates will be different from those evaluated for (conventional) thermalized plasmas.

The present paper is organized as follows; In Section 2, the topology of non-separable ballooning modes [4,5] that can be excited in an axisymmetric confinement configuration is described as geometry plays a key role in the processes identified in later sections. In Section 3, the magnetosonic modes found for multispecies homogeneous plasmas are introduced in order to identify the range of plasma parameters for which the theory of modes emerging in inhomogeneous deuterium – tritium plasmas is developed. In Section 4, the class of mode-particle resonant interactions that are involved in the transfer of energy from the $\alpha$ - particle population to the deuterons are identified. In Section 5, the analysis showing that the considered ballooning modes are localized radially, and their energy is contained, is given. In Section 6, the ballooning profile along magnetic field lines of the considered modes is derived in the absence of mode-particle interactions and, in addition, the (novel) conditions for these interactions are introduced and shown to be intrinsically different from those considered in Section 4, referring to "ordinary" waves. In Section 7, the intrinsically different time and space dependence of the mode amplitude and profile, from that of well-known waves, resulting from resonant interactions with a minority particle population is demonstrated. In Section 8, the amplitudes of the plasma density fluctuations associated with realistic rates of energy extraction from the emitted $\alpha$ - particle population are estimated. In Section 9, results from two different sets of experiments are discussed which lend support to the presented theory. In Section 10, final considerations based on the presented theory are made.

## 2. NON-SEPARABLE HIGH FREQUENCY MODES

We refer, for simplicity, to a toroidal plasma with a large aspect ratio, a circular cross section and high toroidal and poloidal magnetic fields, the former being represented by $B \simeq B_0 / (1 + r\cos\theta / R_0)$ where $R_0$ is the major radius, $r$ the (minor) radial coordinate and $\theta$ the poloidal angle. Moreover, the assumed poloidal field $B_\theta \simeq B_\theta(r)$ is smaller than $B_0$, that is $B_\theta^2 / B_0^2 \ll 1$, and $|dB_\theta / dr| / |B_\theta| \sim |dn/dr| / n \sim 1/a$, $a$ being the torus minor radius. The transverse plasma pressure $p_\perp = (p_e + p_D + p_T)_\perp$ is taken to be $\ll B_0^2 / 8\pi$. Then, to lowest order in the small considered parameters, the radial equilibrium condition reduces to

$$0 = -\frac{\partial}{\partial r}(p_e + p_i)_\perp - \frac{1}{c}(J_\varphi B_\theta - J_\theta B_\varphi). \tag{1}$$

The ballooning modes [4,5] that are introduced for this configuration are represented by plasma density perturbations of the form

$$\hat{n} = \tilde{n}(\theta, r - r_0^0, t) \exp(-i\omega t - im^0 \theta + in^0 \varphi), \tag{2}$$

where $m^0$ and $n^0$ are integers,

$$|\omega| \gg \frac{1}{\tilde{n}} \frac{\partial \tilde{n}}{\partial t}, \tag{3}$$

$$\left|\frac{m^0}{r_0^0}\right| > \left|\frac{1}{\tilde{n}} \frac{\partial \tilde{n}}{\partial r}\right|, \tag{4}$$

$\tilde{n}(\theta, r - r_0^0, t)$ is a non-separable function of $t$, $\theta$, $r - r_0^0$ which is periodic in $\theta$ and is radially localized around the surface $r = r_0^0$, that is with $|(\partial \tilde{n}/\partial r)/\tilde{n}| > 1/a$ where $a$ is the plasma minor radius. Clearly, we are concerned with a special class of ballooning modes. We may then adopt the "disconnected mode" approximation [5], for $|\theta| < \pi$, and reduce $\hat{n}$ to be represented by

$$\hat{n} \simeq \tilde{n}(\theta, t) G(r - r_0^0) \exp\{-i\omega t + in^0[\varphi - q(r - r_0)\theta]\}, \tag{5}$$

where $q(r) \simeq B_0 r / [R_0 B_\theta(r)]$, $q(r = r_0) = m^0 / n^0 \equiv q_0$ and $|r_0 / r_0^0 - 1| < r_0 / R_0$. Then

$$\mathbf{B} \cdot \nabla \hat{n} \simeq \frac{B_0}{q_0 R_0} \frac{\partial \tilde{n}}{\partial \theta} G(r - r_0^0) \exp\{-i\omega t + in^0[\varphi - q(r - r_0)\theta]\}. \tag{6}$$

In particular, we consider $\tilde{n} = \tilde{n}_{ev} + \tilde{n}_{od}$ where $\tilde{n}_{ev}$ and $\tilde{n}_{od}$ are even and odd functions of $\theta$, respectively, $|\tilde{n}_{ev}(\theta = \pi) / \tilde{n}_{ev}(\theta = 0)| \ll 1$, as required for the validity of the disconnected mode approximation, and the component $\tilde{n}_{od}$ is not involved in the cold, homogeneous plasma approximation [6].

## 3. MULTISPECIES MAGNETOSONIC MODES

The modes that, for an infinite, homogeneous and cold plasma correspond to those under consideration can be classified as multispecies magnetosonic modes [6]. These are represented by $\hat{n} = \tilde{n}_k \exp(-i\omega t + ik_\perp y + ik_\parallel z)$, where $k_\perp$ corresponds to $m^0 / r_0$. If we adopt the "disconnected mode" approximation represented by Eq. (5), and refer to Eq. (6), $ik_\parallel \tilde{n}_k$ can simulate $[1/(qR_0)](\partial \tilde{n}/\partial \theta)$.

Since the modes of interest are those that can extract energy [3] from the $\alpha$-particle population produced by the DT fusion reaction, the most appropriate frequency to consider is

$$\omega \simeq \frac{m^0}{r_0} \bar{V}_A = \Omega_\alpha + \delta\omega, \tag{7}$$

where $\bar{V}_A^2 = [B^2/(4\pi n_e m_D)](n_D + 2n_T/3)/n_e$, $\Omega_\alpha$ is the $\alpha$-particle cyclotron frequency and $|\delta\omega| < \Omega_\alpha$. Referring to the dispersion relation presented in Ref. [3], condition (7) corresponds to

$$m^0 \simeq r_0 \omega_{pD}/c \equiv r_0/d_D \gg 1$$

where $\omega_{pD}^2 = 4\pi n_D e^2/m_D$, $d_D = c/\omega_{pD}$, $n_e$, $n_D$ and $n_T$ are the densities of the electron, deuteron and triton populations. In fact, the cold homogeneous plasma dispersion relation {Eq. (6) in Ref. [3]} is

$$\frac{\omega^2 - \bar{\bar{\Omega}}^2}{\omega^2 - \Omega_{Hy}^2}\omega^2 = (k_\perp^2 + k_\parallel^2)\bar{V}_A^2, \tag{8}$$

where $\bar{\bar{\Omega}} = (n_D\Omega_T + n_T\Omega_D)/n$, $\Omega_{Hy}^2 = \Omega_D\Omega_T(\bar{\bar{\Omega}}/\bar{\Omega})$ and $\bar{\Omega} = (n_D\Omega_D + n_T\Omega_T)/n$. We can verify that for $\omega \simeq \Omega_D \simeq 5 \times 10^8 (B/10\,T)\,rad/s$, and $n_D = n_T = n/2$, $\omega^2$ is close to $(k_\perp^2 + k_\parallel^2)\bar{V}_A^2$.

## 4. "CONVENTIONAL" MODE-PARTICLE RESONANT INTERACTIONS

Referring, for simplicity, to the homogeneous model, we note that the mode-particle resonance $\omega - \Omega_\alpha + k_\parallel v_\parallel)_\alpha = 0$ is involved in extracting energy [3] from the $\alpha$-particle population. For this, significant values of $k_\perp \rho_\alpha$ have to be considered [1], where $\rho_\alpha = V_\alpha/\Omega_\alpha$, $V_\alpha$ being the velocity of the emitted $\alpha$-particles. We note that, considering $n_T \simeq n_D$, the value of $k_\perp d_D$ is the main parameter that identifies the relevant limits of the dispersion relation. The ratio $\rho_\alpha/d_D = V_\alpha/V_{AD}$ is computed where $V_{AD}^2 = B^2/(4\pi m_D n_D)$. In particular, $k_\perp \rho_\alpha \simeq (3.5 n_D/10^{21} m^{-3})^{1/2}(10\,T/B)k_\perp d_D$ and $k_\perp d_D \sim 1$ can correspond to significant value of $k_\perp \rho_\alpha$ for attainable plasma confinement parameters. Clearly, when referring to the ballooning modes represented by Eq. (5) $m^0/r_0$ corresponds to $k_\perp$.

Referring to the energy absorbing resonance $\omega - \Omega_D + k_\parallel v_\parallel)_D = 0$ we notice that $k_\parallel v_\parallel)_D \simeq k_\parallel v_\parallel)_\alpha$ and the needed two resonances with the two populations will have to involve different $k_\parallel$'s. Moreover, mode particle resonant interactions with the main body of the electron distribution, transferring considerable energy to it, are avoided as $\bar{\omega}_{te} \ll \Omega_\alpha \ll \Omega_e$ where $\bar{\omega}_{te}$ is the average electron transit frequency. On the other hand, the transfer of energy to the tail of the electron distribution corresponding to $\Omega_\alpha = k_\parallel v_\parallel)_e$ should be considered as a plasma diagnostic means [8] involving the emission of e.m. radiation at the frequency $\omega = \Omega_\alpha$. This has been, in fact, observed by the DT plasma experiments reported in Ref. [9]. Moreover, since finite values of $k_\perp \rho_\alpha$ are involved, higher harmonic modes can be excited as well. In the case of the deuterons we have to consider that, correspondingly, $(k_\perp \rho_D)^2 \ll 1$.

## 5. RADIAL CONTAINMENT

The radial localization of the modes we are considering is like that of the waves which were proposed [9] as being responsible for the experimentally observed e. m. emission at the harmonics of $\alpha$ - cyclotron frequency. The radial containment of a mode can, in fact, be viewed as a positive characteristic in view of preventing loss of its energy toward surrounding walls.

As indicated in the previous section we choose to consider mode spatial profiles represented by $\tilde{n}(r - r_0, \theta) = \tilde{g}(r - r_0^0)\tilde{\tilde{f}}(\theta, r_0^0)$. In order to evaluate $\tilde{\tilde{g}}(r - r_0^0)$ we neglect the effects of toroidicity that are included in $\tilde{\tilde{f}}(\theta, r_0^0)$. We refer to modes with frequencies $\omega \simeq m^0 V_A^0 / r_0$ where $V_A^0 = B / [4\pi \rho(r_0)]^{1/2}$, $\rho = \bar{m}_i n(r) = m_D n_D + m_T n_T$, and $(m^0)^2 \gg 1$.

Then, adopting the ideal MHD approximation and following a standard procedure {see Ref. [8]}, based on the same equations considered in Section 6, we are led to find the following equation for $\tilde{\tilde{g}}(r - r_0^0)$

$$\frac{d^2 \tilde{\tilde{g}}}{dr^2} - \left[ \frac{(m^0)^2}{r^2} - \frac{\omega^2}{(V_A^0)^2} \frac{n(r)}{n_0} \right] \tilde{\tilde{g}} = 0 \tag{9}$$

where $n_0$ is the peak particle density and $(V_A^0)^2 = B^2 / (4\pi n_0 m_i)$. Then $\omega_0^2 / (m^0)^2$ can be chosen in such a way that $r_0^0$ defined by

$$\frac{(m^0)^2 (V_{A0}^0)^2}{r_0^0 \omega_0^2} = 1, \tag{10}$$

for $(V_{A0}^0)^2 = B^2 / [4\pi m_i n(r = r_0^0)]$, and by

$$2 \frac{(m^0)^2}{(r_0^0)^3} + \frac{\omega_0^2}{(V_{A0}^0)^2} \left( \frac{1}{n} \frac{dn}{dr} \right)_{r=r_0^0} = 0 \tag{11}$$

falls within the plasma column. In fact, Eqs. (10) and (11) imply that

$$\left( \frac{1}{n} \frac{dn}{dr} \right)_{r=r_0^0} = -\frac{2}{r_0^0}. \tag{12}$$

Consequently, Eq. (9) reduces to

$$\frac{d^2\tilde{\tilde{g}}}{dr^2} - \left\{\left[\frac{6(m^0)^2}{(r_0^0)^4} - \frac{\omega_0^2}{(V_A^0)^2}\frac{1}{n_0}\frac{d^2 n}{dr^2}\right]\frac{1}{2}(r-r_0^0)^2 - \frac{(\delta\omega_1^2)}{(V_A^0)^2}\frac{n(r_0^0)}{n_0}\right\}\tilde{\tilde{g}} = 0 \qquad (13)$$

for $\omega^2 \simeq \omega_0^2 + (\delta\omega^2)_1$. The relevant solution is

$$\tilde{\tilde{g}} = \tilde{\tilde{g}}_0 \exp\left[-\frac{1}{2}\frac{(r-r_0^0)^2}{\Delta_r^2}\right] \qquad (14)$$

where

$$\Delta_r^4 = \frac{(r_0^0)^4}{3(m^0)^2} \bigg/ \left[1 - \frac{(r_0^0)^2}{6}\left(\frac{1}{n}\frac{d^2 n}{dr^2}\right)_{r=r_0^0}\right] \ll a^4, \qquad (15)$$

$a$ is the plasma minor radius and

$$(\delta\omega^2)_1 \simeq \omega_0^2 \Delta_r^2 \left(\frac{1}{n}\frac{dn}{dr}\right)_{r=r_0^0}. \qquad (16)$$

In particular,

$$\frac{(\delta\omega^2)_1}{\omega_0^2} = \left(\frac{r_0^0}{m^0 \Delta_r}\right)^2$$

implies that $(m^0)^2 > (r_0^0/\Delta_r)^2$ as assumed. As an example, we note that if $n \simeq n_0(1-r^2/a^2)$, condition corresponds to $r_0^0 = a/\sqrt{2}$.

## 6. INITIAL BALLOONING MODE PROFILE

The derivation of the ballooning mode equation is greatly simplified by the observation made in Section 3 that modes with $\omega \simeq \Omega_D$ and $kd_D \sim 1$ can be described by the relatively simple theory of magnetosonic modes where both deuterons and tritons can be treated in the limit $\omega^2 > \Omega_D^2$. Then the particle conservation equation

$$-i\omega\hat{n} + n\nabla \cdot \hat{\mathbf{u}}_e \simeq 0 \qquad (17)$$

is combined with the total momentum conservation equation

$$-i\omega(m_D n_D \hat{\mathbf{u}}_D + m_T n_T \hat{\mathbf{u}}_T) = -\nabla\left(\hat{p}_e + \hat{p}_i + \frac{\hat{\mathbf{B}}\cdot\mathbf{B}}{4\pi}\right) + \frac{1}{4\pi}\left(\mathbf{B}\cdot\nabla\hat{\mathbf{B}} + \hat{\mathbf{B}}\cdot\nabla\mathbf{B}\right) \qquad (18)$$

where the contribution of the $\alpha$-particle population, considered a minority species, is not included and $\hat{n} = \hat{n}_D + \hat{n}_T = \hat{n}_e$.

Referring to the toroidal configuration introduced in Section 2, we take $B \simeq B_m + \Delta B$, where $B_m$ is the minimum field corresponding to $\theta = 0$, $B_m \simeq B_0(1-r/R_0)$. Then $\Delta B/B_0 \simeq (r/R_0)(1-\cos\theta)$. The equation $\hat{\mathbf{E}} + \hat{\mathbf{u}}_e \times \mathbf{B}/c = 0$ combined with $-\partial\hat{\mathbf{B}}/\partial t = c\nabla\times\hat{\mathbf{E}}$ leads to

$$-i\omega\hat{\mathbf{B}} = \mathbf{B}\cdot\nabla\hat{\mathbf{u}}_e - \mathbf{B}(\nabla\cdot\hat{\mathbf{u}}_e) \tag{19}$$

that is

$$-i\omega\hat{B} = B(\nabla_{\|}\hat{u}_{e\|} - \nabla\cdot\hat{\mathbf{u}}_e),$$

where $\nabla_{\|} = (\mathbf{B}/B)\cdot\nabla$, and

$$\frac{\hat{B}}{B} = -\frac{i}{\omega}(\nabla\cdot\hat{\mathbf{u}}_e - \nabla_{\|}\hat{u}_{e\|}). \tag{20}$$

Therefore,

$$\frac{\hat{B}}{B} = \frac{\hat{n}}{n} + \frac{i}{\omega}(\nabla_{\|}\hat{u}_{e\|}). \tag{21}$$

Next, considering that $\beta \equiv 8\pi(p_e + p_i)/B^2 \ll 1$ we take $|\hat{p}_e| \sim |\hat{p}_i| \ll |\hat{\mathbf{B}}\cdot\mathbf{B}/4\pi|$ and, for $\hat{\rho} = \hat{n}m_D(n_D + m_D n_T/m_T)/n$, obtain

$$\omega^2 \hat{\rho} \simeq -\frac{1}{4\pi}\nabla^2(\hat{\mathbf{B}}\cdot\mathbf{B}). \tag{22}$$

Combining this with Eq. (21) we arrive at the mode dispersion equation indicating that $\omega^2 = (V_{Am}m^0/r_0)^2 + (\delta\omega^2)_1 + (\delta\omega^2)_2$, where $(\delta\omega^2)_2$ can be evaluated referring, for simplicity, to the surface $r = r_0 \simeq r_0^0$ from the solution of the following ballooning equation

$$(\delta\omega^2)_2 \tilde{\rho}(r_0,\theta) \simeq -\frac{\bar{V}_{Am}^2}{(qR_0)^2}\left(\frac{d^2\tilde{\rho}}{d\theta^2}\right) - V_{Am}^2\left[2\left(\frac{m^0}{r_0}\right)^2 \frac{r_0}{R_0}(1-\cos\theta)\right]\tilde{\rho}(r_0,\theta) \simeq 0. \tag{23}$$

This indicates that the mode is localized over a relatively small angle $\Delta\theta$, around $\theta = 0$, that is

$$\Delta\theta \sim \frac{1}{(qm^0)^{1/2}}\left(\frac{r_0}{R_0}\right)^{1/4} < 1. \tag{24}$$

Then the solution of Eq. (23) is

$$\tilde{\rho}(r_0,\theta) \simeq \tilde{\rho}(r_0)\exp\left(-\sigma\frac{\theta^2}{2}\right) \tag{25}$$

where

$$\sigma^2 = (m^0 q)^2 \frac{R_0}{r_0} \tag{26}$$

and

$$(\delta\omega^2)_2 = \left(\frac{V_{Am}}{qR_0}\right)^2. \tag{27}$$

Consequently, the distance along a magnetic field line over which the mode is localized is

$$L_{\|} \sim qR_0 \frac{\Delta\theta}{2\pi}. \tag{28}$$

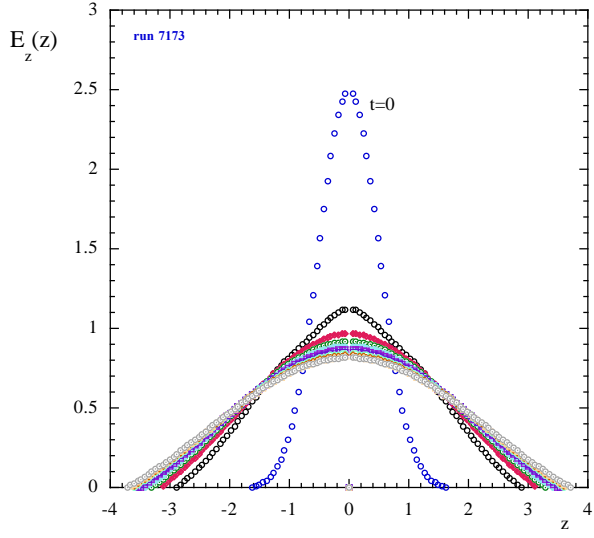

*FIG.1 Ballooning mode evolution resulting from (damping) mode-particle resonances with a Maxwellian distribution. The considered ballooning mode profile is represented by Eq. (32). Courtesy of A. Cardinali]*

Assuming that the $\alpha$-particle density is relatively small we refer to Eq. (22) and extend it by adding the contribution of $\hat{\mathbf{P}}_\alpha$, the perturbed $\alpha$-particle pressure tensor. Clearly, this can be derived from the distribution function $\hat{f}_\alpha$ based on an unperturbed $f_\alpha\left(r_0, v_\perp^2, v_\parallel\right)$. Considering the symmetry in $\theta$ of the driving factors of the perturbed $\hat{f}_\alpha$ (see Appendix), this can be split into an even component of $\theta$, $\hat{f}_{\alpha-even}$, and an odd component $\hat{f}_{\alpha-odd}$. Then the mode-particle resonance condition is [see Eq. (A-9)]

$$0 = \left[\left(\delta\omega_\alpha\right) - \frac{r_0}{2R_0}\theta^2\Omega_\alpha\right] + v_\parallel^2 \frac{1}{\left(q_0 R_0\right)^2} \frac{1}{G_\alpha} \frac{dG_\alpha}{d\theta} \left\{\left[\left(\delta\omega_\alpha\right) - \frac{r_0}{2R_0}\theta^2\Omega_\alpha\right]^{-1} \frac{dG_\alpha}{d\theta}\right\}, \quad (29)$$

where $G_\alpha\left(\theta^2\right)$ represents the ballooning profile of $\hat{f}_{\alpha-even}$. As an example, if $G_\alpha = \exp\left[-\theta^2/\left(2|\Delta\theta|^2\right)\right]$,

$$\frac{1}{G_\alpha}\frac{d^2 G_\alpha}{d\theta^2} = -\frac{1}{|\Delta\theta|^2}\left(1 - \frac{\theta^2}{|\Delta\theta|^2}\right).$$

For $|\delta\omega_\alpha| \gg \left(r_0/R_0\right)\Omega_\alpha\theta^2$ and remembering that $|k_\parallel|$ can be represented by $|\theta|/\left[|\Delta\theta|^2 q_0 R_0\right]$, we would have, instead of Eq. (29),

$$\left(\delta\omega_\alpha\right)^2 - v_\parallel^2\left(|k_\parallel|_{max}^2 - |k_\parallel|^2\right) = 0. \quad (30)$$

Then, if we define

$$\left(\delta V_{ph}\right) \equiv \left(\delta\omega_\alpha\right)\left(q_0 R_0 |\Delta\theta|\right),$$

the resonant $v_{\|}$ is given by

$$v_{\|}^2\big)_{\text{Res}} = \left(\delta V_{ph}\right)^2 \frac{1}{1-\theta^2/|\Delta\theta|^2}. \tag{31}$$

Considering the mode-particle resonance conditions represented by Eq. (29) it is clear that the most effective damping of the mode should correspond to deuteron velocities that are as close as possible to $V_\alpha$ but not so close that the resonating particle densities are too small, as in the case of particles far out in the tail of a Maxwellian distribution. Thus, a simple mathematical model incorporating these two requirements indicates that the resonant velocities, for most effective damping, corresponds to a fraction of $V_\alpha$ that is in the tail of the deuteron distribution. This is consistent with the experimental observation reported in Ref [10].

## 7. EVOLUTION OF BALLOONING MODES

Since the considered mode is contained (standing and localized) in the radial direction [8], this feature should prevent it from transporting its energy toward the wall surrounding the plasma. The other important feature is that the mode is of the ballooning type along the magnetic field and as such it can be viewed as a superposition of modes with the same frequency but propagating along the field with different phase velocities [2]. In this case the relevant mode-particle interactions that can produce damping or growth of the mode can affect the height and the width (along the magnetic field) of the mode ballooning amplitude.

Referring to the combination of growth and damping resulting from the interaction of the considered modes with the reaction products and the fusing nuclei, we may expect that the modes will evolve to become purely oscillatory where the growth and damping rates compensate each other. We expect also that the mode radial profile will change during its evolution as shown by the following analysis referring to the case where damping prevails. An oscillatory ballooning mode viewed as a superposition of standing modes having the same frequencies and involving a continuous spectrum of the relevant phase velocities [2], can be represented by

$$\hat{n} \propto \exp\left[-\frac{\theta^2}{2(\Delta\theta)^2} - i\omega t\right] \propto \exp\left[-\frac{l^2}{2(\Delta l)^2} - i\omega t\right], \tag{32}$$

where $(\Delta\theta)^2 < 1$ and $l = R_0 q_0 \theta$. Thus, the superposed waves have the form

$$\exp\left[-2k_l^2(\Delta l)^2 - i(\omega t - k_l l)\right]. \tag{33}$$

If we consider the case analyzed in Section 6, the relevant mode-particle resonance condition, for $|\delta\omega_\alpha| \gg (r_0/R_0)|\Delta\theta|^2 \Omega_\alpha$, is

$$\left(\delta\omega_\alpha\right)^2 - \left(k_l v_{\|}\right)^2 = 0. \tag{34}$$

Returning to the simpler case where $\omega = \omega_0 - i\gamma_d(k_l^2)$ and $\gamma_d > 0$, if the following model for the damping rate corresponding to all values of $k_l$ is assumed

$$\gamma_d\left(k_l^2\right) = 2\bar{\gamma} k_l^2 \left(\Delta l\right)^2, \tag{35}$$

we find

$$\hat{n} \propto \frac{1}{(\Delta l)(1+\bar{\gamma}t)^{1/2}} \exp\left[-\frac{l^2}{2(\Delta l)^2(1+\bar{\gamma}t)} - i\omega_0 t\right]. \tag{36}$$

Clearly, this indicates that the resulting mode profile becomes broadened and lowered as $t$ increases. In fact, Eq. (35) can be considered the limit of a less simplified expression such as

$$\gamma_d\left(k_l^2\right) = 2\bar{\gamma}\left(\Delta l\right)^2 \frac{k_l^2}{1+k_l^4/k_M^4}, \tag{37}$$

for $k_l^4 \ll k_M^4$.

Starting from an initial $(t=0)$ mode profile represented by Eq. (32) an accurate numerical analysis (see Fig. 1) of the mode profile evolution has been carried out involving the relevant Landau damping with a Maxwellian distribution. This has involved decomposing $\tilde{n}(l) = \tilde{n}_0 \exp\left\{-l^2/\left[2(\Delta l)^2\right]\right\}$ into Fourier harmonics as shown by Eq. (33). For each harmonic the damping associated with the mode-particle resonance condition (34) has been evaluated and the resulting time evolution of the same harmonic has been followed. Then the ballooning mode has been recomposed as a function of time with the results shown in Fig. 1. In this case the features of the Maxwellian distribution have not made it necessary to restrict $\gamma_d\left(k_l^2\right)$ to the model represented by Eq. (35).

Since the process of transferring energy from populations with high energy to populations with lower energies through the excitation of ballooning modes can avoid the inefficiencies of conventional nonlinear coupling processes [3], further numerical analysis involving resonances producing both growth and damping on components of the same ballooning mode is planned. Clearly, the indications of the analysis in Section 6 will have to be considered.

## 8. PLASMA FLUCTUATIONS AND POWER DENSITIES

An important question is whether the considered modes will involve acceptable plasma density fluctuations in order to transfer energy at significant rates from the $\alpha$ - particle population to the reacting deuterons.

We may consider $Q_\alpha^0 = 1 \text{ MW}/m^3 = J/(cm^3 s)$ as a reference value for the power density of the emitted $\alpha$ - particles and we assume that this is also a reference value for the power density of the modes driven by their interaction with the emitted $\alpha$ - particles. Then, referring to a homogeneous plasma model, the corresponding mode amplitudes could be evaluated from

$$Q_\alpha \simeq \gamma_\alpha \varepsilon_W$$

where $\varepsilon_W$ is the energy density of the modes and $\gamma_\alpha$ is the rate of energy extraction from the $\alpha$ - particle population, and

$$\gamma_\alpha \varepsilon_W = \gamma_\alpha \frac{\left|\hat{B}_k\right|^2}{4\pi} = 2\gamma_\alpha \left|\frac{\hat{B}_k}{B}\right|^2 \frac{B^2}{8\pi}. \tag{38}$$

Therefore, the plasma density fluctuations associated with the extracted power density $Q_\alpha$ can be estimated as

$$\left|\frac{\hat{n}_k}{n}\right| \simeq \left|\frac{\hat{B}_k}{B}\right| \simeq \left(\frac{\bar{Q}_\alpha}{2\gamma_\alpha}\right)^{1/2} \left(\frac{1.6\,T}{B}\right), \tag{39}$$

where $\bar{Q}_\alpha \equiv Q_\alpha / Q_\alpha^0$, $|\hat{n}_k|/n \ll 1$ and $\gamma_\alpha \ll \Omega_\alpha \simeq 2 \times 10^8 (B/10\,T)\,rad/s$. Thus, if $\bar{Q}_\alpha \simeq 1$, $B \simeq 10\,T$ and $2\gamma_\alpha \simeq 10^4\,s$, $|\hat{n}_k/n| \simeq 1.6 \times 10^{-3}$ that is a modest value. Clearly, an estimate of $\gamma_\alpha$, associated with the resonance (29), will depend on the details of the considered $\alpha$ - particle distribution. The advantage of adopting high magnetic fields, for a fixed value of $\bar{Q}_\alpha / \gamma_\alpha$, is clear from Eq. (39). When all the energy extracted from the $\alpha$ - population is transferred to the deuterons we have

$$Q_\alpha = P_{abs} = \gamma_D \varepsilon_W \tag{40}$$

where $\gamma_D$ is the relevant damping rate.

As a start, we may assume a Maxwellian distribution for the deuterons and consider $k_\perp^2 \rho_D^2 \ll 1$ where $\rho_D^2 = 2T_D / (m_D \Omega_D^2)$ for the relevant perturbations. Given the characteristics of the mode-particle resonant interactions discussed earlier, one of the difficulty of relying on a homogeneous plasma model, to have an estimate of $\gamma_D$, is that the result depends on a significant choice for a representative value of $k_\parallel$. In fact, the mode-particle resonances represented by Eq. (29) depends on $\theta$ and the consequence of this is that the mode profile, along the magnetic field, can change as a function of time as shown by the analysis of Section 7. Therefore, the definitions of $\gamma_\alpha$ and $\gamma_D$ will have to be reformulated accordingly. In particular, referring for simplicity to Eq. (30) and considering that superthermal deuterons should be involved in absorbing the mode energy, we may take

$$(\delta V_{ph})^2 \simeq \alpha_{ST} (2T_D / m_D) \tag{41}$$

with $\alpha_{ST} \geq 1$. In this case the corresponding mode amplitude is its maximum, as $v_\parallel\big)_{Res}^D = (\delta V_{ph})$, is reached at $\theta = 0$. On the other hand, if $V_\alpha^2 \gg (\delta V_{ph})^2$ a significant resonance with the $\alpha$ - particle population is reached for $\theta^2$ near $|\Delta\theta|^2$.

Finally, we observe that when considering higher harmonics of $\Omega_\alpha$ the mode-particle resonance condition (29) is replaced by

$$(\omega - p^0 \Omega_\alpha)^2 - v_\parallel^2 \frac{1}{(q_0 R_0)^2 |\Delta\theta|^2}\left(1 - \frac{\theta^2}{|\Delta\theta|^2}\right) = 0. \tag{42}$$

Clearly, the coupling of higher harmonics with the lowest harmonic $(p^0 = 1)$ deserves to be taken into consideration in view of the features of the perturbed deuteron density and for $(k_\perp \rho_D)^2 \ll 1$.

## 9. RELEVANT EXPERIMENTAL OBSERVATIONS

Results that are relevant to the theory described in the previous sections have been obtained by a series of experiments [10] carried out on magnetically confined plasmas with a combined mirror – FRC confinement configuration. These involved Deuterium plasmas with relatively low temperatures in which a Hydrogen neutral beam with relatively high energy (30 keV) was injected. A large increase of neutron emission, due to DD reactions, was observed. This is clearly due to a collective mode driven by high energy protons and supplying energy to the tail of the (energy) distribution of the reacting nuclei. A coherent mode with the Deuteron cyclotron frequency was observed at the same time. Differently from the theory involving DT plasmas, the cyclotron frequency of the high energy population is larger than that of the reacting nuclei. On the other hand, considering that most of the area of the transverse cross section of the plasma column is subject to an axisymmetric magnetic field distribution of the mirror kind, the emergence of a ballooning mode [11] of the kind analyzed in the previous section can be envisioned. In this case the magnetic field near $z = 0$, $z$ being the symmetry axis would be of the form $B \simeq B_0 \left( 1 + l^2 / L^2 \right)$, $l$ being the distance along a given field line. Then the analysis summarized in Section 7 can be extended to cover this case. Clearly, it would be desirable to conduct parallel experiments on plasmas with different magnetic confinement configurations.

The first set of experiments with DT plasmas carried out by the JET facility [9] revealed significant rate of e.m. radiation emission from the lowest to high harmonics of $\Omega_\alpha$. Assuming that modes of the kind analyzed here were excited, a coupling to e.m. modes with observed frequency and propagating away from the plasma column should be considered. Since neutrons of non-thermal origin have also been observed it is conceivable that a fraction of them could be associated with the excitation of relevant collective modes.

10. FINAL CONSIDERATIONS

Therefore, by identifying and possibly controlling the modes that can increase the reaction rates, for a given temperature of the electron population, can lead to new perspectives that include D-T ignition under less restrictive conditions than those usually assumed, utilizing high magnetic field experiments to reach significant burn conditions with D-D catalyzed reaction, etc. On the other hand, new experiments will be needed to ascertain the excitation of the considered modes – an objective that can be pursued even with plasmas that do not contain tritium, as indicated earlier. Another issue that needs consideration is that of gaining some control on the process by which the amplitudes of these modes are limited. A damping on a region of the distribution in phase space of the reacting nuclei which can maximize the plasma reactivity is certainly desirable. In this context, the technology developed, and the expertise gained in introducing of Ion Cyclotron Resonant Heating (ICRH) systems, may make it possible to interact from outside the plasma column with the considered modes without the high-power requirements associated with "conventional" ion cyclotron heating.

**ACKNOWLEDGEMENTS**

We are indebted to A. Cardinali and V. Ricci for their contributions to the first phase of this ongoing work and for their continuing concern. Sponsored in part by the Kavli Foundation (through MIT) and by CNR (Consiglio Nazionale delle Ricerche) of Italy.

## APPENDIX

Driven Distribution Functions

Referring at first to the $\alpha$ - particle population we start from the linearized collisionless equation

$$\left[\frac{\partial}{\partial t} + \mathbf{v}\cdot\nabla + \frac{q_\alpha}{m_\alpha c}(\mathbf{v}\times\mathbf{B})\cdot\frac{\partial}{\partial \mathbf{v}}\right]\hat{f}_\alpha(\mathbf{r},\mathbf{v},t) + \frac{q_\alpha}{m_\alpha}\left(\hat{\mathbf{E}} + \frac{\mathbf{v}\times\hat{\mathbf{B}}}{c}\right)\cdot\frac{\partial}{\partial \mathbf{v}}f_\alpha(\mathbf{v}) = 0, \qquad (A-1)$$

and refer to the coordinates $v_\parallel$ and $v_\perp$ in velocity space, where $v_\parallel \equiv \mathbf{v}\cdot\mathbf{B}/B$, $\mathbf{v}_\perp \equiv \mathbf{v} - v_\parallel \mathbf{B}/B$, $v_r = v_\perp \cos\bar{\varphi}$, $v_\theta = v_\perp \sin\bar{\varphi}$, $v_\perp = \sqrt{v_r^2 + v_\theta^2}$ and $\bar{\varphi} = \tan^{-1}(v_\theta/v_r)$. Then

$$\mathbf{v}\cdot\nabla\hat{f}_\alpha = i\frac{m^0}{r_0}v_\perp \sin\bar{\varphi}\,\hat{f}_\alpha + v_\parallel \frac{\partial}{\partial l}\hat{f}_\alpha$$

$$\frac{q_\alpha}{m_\alpha c}(\mathbf{v}\times\mathbf{B})\cdot\frac{\partial}{\partial \mathbf{v}}\hat{f}_\alpha = -\Omega_\alpha \frac{\partial}{\partial \bar{\varphi}}\hat{f}_\alpha$$

$$\hat{\mathbf{E}} \cdot \frac{\partial}{\partial \mathbf{v}} f_\alpha(\mathrm{v}_\perp, \mathrm{v}_\parallel) = 2\hat{\mathbf{E}}_\perp \cdot \mathbf{v}_\perp \frac{\partial}{\partial \mathrm{v}_\perp^2} f_\alpha + \hat{E}_\parallel \frac{\partial}{\partial \mathrm{v}_\parallel} f_\alpha$$

where $\partial/\partial l \simeq [1/(q_0 R_0)] \partial/\partial \theta$. Moreover

$$(\mathbf{v} \times \hat{\mathbf{B}}) \cdot \frac{\partial}{\partial \mathbf{v}} f_\alpha(\mathrm{v}_\perp, \mathrm{v}_\parallel) = 0$$

if we assume for simplicity that $f_\alpha$ is isotropic. Consequently, Eq. (A-1) reduces to

$$-i\omega \hat{f}_\alpha + i\frac{m^0}{r_0} \mathrm{v}_\perp \sin\bar{\varphi} \hat{f}_\alpha + \mathrm{v}_\parallel \frac{\partial}{\partial l} \hat{f}_\alpha - \Omega_\alpha \frac{\partial}{\partial \bar{\varphi}} \hat{f}_\alpha$$
$$= -\frac{q_\alpha}{m_\alpha}\left\{(\hat{E}_r \cos\bar{\varphi} + \hat{E}_\theta \sin\bar{\varphi})\frac{\partial}{\partial \mathrm{v}_\perp} f_\alpha + \hat{E}_\parallel \frac{\partial}{\partial \mathrm{v}_\parallel} f_\alpha\right\}.$$

(A-2)

We define

$$\hat{f}_\alpha(l, \mathrm{v}_\parallel, \mathrm{v}_\perp, \bar{\varphi}) \equiv \hat{g}_\alpha(l, \mathrm{v}_\parallel, \mathrm{v}_\perp, \bar{\varphi}) \exp(-i\mu_\alpha \cos\bar{\varphi})$$

(A-3)

where $\mu_\alpha \equiv (m^0/r_0)(\mathrm{v}_\perp/\Omega_\alpha)$ and find

$$\left(-i\omega + \mathrm{v}_\parallel \frac{\partial}{\partial l}\right)\hat{g}_\alpha - \Omega_\alpha \frac{\partial}{\partial \bar{\varphi}} \hat{g}_\alpha$$
$$= -\frac{q_\alpha}{m_\alpha}\exp(i\mu_\alpha \cos\bar{\varphi})\left\{(\hat{E}_r \cos\bar{\varphi} + \hat{E}_\theta \sin\bar{\varphi})\frac{\partial}{\partial \mathrm{v}_\perp} f_\alpha + \hat{E}_\parallel \frac{\partial}{\partial \mathrm{v}_\parallel} f_\alpha\right\}$$

(A-4)

Next, use $\hat{g}_\alpha(l, \mathrm{v}_\parallel, \mathrm{v}_\perp, \bar{\varphi}) = \sum_{n=-\infty}^{+\infty} \hat{g}_{\alpha,n}(l, \mathrm{v}_\parallel, \mathrm{v}_\perp) \exp(-in\bar{\varphi})$ in Eq. (A-4) and obtain

$$\sum_{n=-\infty}^{+\infty}\left(-i\omega + in\Omega_\alpha + \mathrm{v}_\parallel \frac{\partial}{\partial l}\right)\hat{g}_{\alpha,n}(l, \mathrm{v}_\parallel, \mathrm{v}_\perp)\exp(-in\bar{\varphi})$$
$$= -\frac{q_\alpha}{m_\alpha}\exp(i\mu_\alpha \cos\bar{\varphi})\left\{(\hat{E}_r \cos\bar{\varphi} + \hat{E}_\theta \sin\bar{\varphi})\frac{\partial}{\partial \mathrm{v}_\perp} f_\alpha + \hat{E}_\parallel \frac{\partial}{\partial \mathrm{v}_\parallel} f_\alpha\right\}$$

(A-5)

Now we use the following expressions

$$\sum_{n=-\infty}^{+\infty} i^n J_n(\mu_\alpha) \exp(-in\bar{\varphi}) = \exp(i\mu_\alpha \cos\bar{\varphi}),$$
$$\sum_{n=-\infty}^{+\infty} i^n n J_n(\mu_\alpha) \exp(-in\bar{\varphi}) = \mu_\alpha \sin\bar{\varphi} \exp(i\mu_\alpha \cos\bar{\varphi}),$$
$$\sum_{n=-\infty}^{+\infty} i^n J'_n(\mu_\alpha) \exp(-in\bar{\varphi}) = i\cos\bar{\varphi} \exp(i\mu_\alpha \cos\bar{\varphi}),$$

(A-6)

and obtain from Eq. (A-5),

$$\sum_{n=-\infty}^{+\infty}\left(-i\omega + in\Omega_\alpha + \mathrm{v}_\parallel \frac{\partial}{\partial l}\right)\hat{g}_{\alpha,n} \exp(-in\bar{\varphi})$$
$$= i\frac{q_\alpha}{m_\alpha}\sum_{n=-\infty}^{+\infty} i^n\left\{\left[J'_n(\mu_\alpha)\hat{E}_r + i\frac{n}{\mu_\alpha} J_n(\mu_\alpha)\hat{E}_\theta\right]\frac{\partial}{\partial \mathrm{v}_\perp} f_\alpha + iJ_n(\mu_\alpha)\hat{E}_\parallel \frac{\partial}{\partial \mathrm{v}_\parallel} f_\alpha\right\}\exp(-in\bar{\varphi}).$$

(A-7)

Thus

$$\left(-i\omega + in\Omega_\alpha + v_\parallel \frac{\partial}{\partial l}\right)\hat{\tilde{g}}_{\alpha,n}\left(l, v_\parallel, v_\perp, \mathbf{r}_\perp\right)$$
$$= \frac{q_\alpha}{m_\alpha} i^{n+1} \left\{\left[J'_n(\mu_\alpha)\hat{E}_r + i\frac{n}{\mu_\alpha}J_n(\mu_\alpha)\hat{E}_\theta\right]\frac{\partial}{\partial v_\perp}f_\alpha + iJ_n(\mu_\alpha)\hat{E}_\parallel \frac{\partial}{\partial v_\parallel}f_\alpha\right\}. \quad (A\text{-}8)$$

In particular for $n=1$ and $\omega = \Omega_\alpha + \delta\omega_\alpha$

$$\left(-i\delta\omega_\alpha + v_\parallel \frac{\partial}{\partial l}\right)\tilde{g}_{\alpha,1}\left(l, v_\parallel, v_\perp\right)$$
$$= -\frac{q_\alpha}{m_\alpha}\left\{\left[J'_1(\mu_\alpha)\tilde{E}_r + i\frac{1}{\mu_\alpha}J_1(\mu_\alpha)\tilde{E}_\theta\right]\frac{\partial}{\partial v_\perp}f_\alpha + iJ_1(\mu_\alpha)\tilde{E}_\parallel \frac{\partial}{\partial v_\parallel}f_\alpha\right\}. \quad (A\text{-}9)$$

Now if we separate $\tilde{g}_{\alpha,1}$ into an even and odd functions and take into account the parity of the r.h.s. of Eq. (A-8) we arrive at the mode-particle resonance (8-4).

Referring to Eq. (A-3) we have

$$\hat{f}_\alpha\left(l, v_\parallel, v_\perp, \bar{\varphi}\right) = \sum_{n=-\infty}^{+\infty} \hat{\tilde{g}}_{\alpha,n}\left(l, v_\parallel, v_\perp\right)\exp\left(-in\bar{\varphi} - i\mu_\alpha \cos\bar{\varphi}\right). \quad (A\text{-}10)$$

When the expression

$$\exp\left(-i\mu_\alpha \cos\bar{\varphi}\right) = \sum_{m=-\infty}^{+\infty} (-i)^m J_m(\mu_\alpha)\exp(im\bar{\varphi}) \quad (A\text{-}11)$$

is used, Eq. (A-10) gives

$$\int d\bar{\varphi}\, \hat{f}_\alpha\left(l, v_\parallel, v_\perp, \bar{\varphi}\right) = 2\pi \sum_{n=-\infty}^{+\infty} (-i)^n J_n(\mu_\alpha)\hat{\tilde{g}}_{\alpha,n}\left(l, v_\parallel, v_\perp\right), \quad (A\text{-}12)$$

where $\hat{\tilde{g}}_{\alpha,n}\left(l, v_\parallel, v_\perp\right)$ is determined by Eq. (A-8).

For the modes that we consider we may take $\hat{E}_\varphi = 0$, $\hat{\mathbf{E}}_\perp \simeq \hat{E}_r \mathbf{e}_r$ and $(\omega/c)\hat{B}_\varphi \simeq -i(m^0/r_0)\hat{E}_r$. Since $\hat{B}_\varphi/B \propto \hat{n}/n$, $\hat{B}_\varphi$ and $\hat{E}_r$ are even functions of $\theta$.

To assess the effect of the mode-particle resonance for the deuteron population we consider, for simplicity, the homogeneous plasma model. In this case, we refer to Eq. (A-8), where we take $\tilde{E}_z = 0$, $\partial/\partial l = ik_\parallel$, and find $\tilde{f}_D$, the perturbed deuteron distribution function, as

$$\tilde{f}_D = -\frac{e}{m_D}\sum_{n=-\infty}^{+\infty} i^n \frac{\exp(-in\bar{\varphi} - i\mu_D \cos\bar{\varphi})}{\omega - n\Omega_D - k_\parallel v_\parallel}\left[J'_n(\mu_D)\tilde{E}_x + i\frac{n}{\mu_D}J_n(\mu_D)\tilde{E}_y\right]\frac{\partial}{\partial v_\perp}f_D^M \quad (A\text{-}13)$$

where $\mu_D \equiv k_y v_\perp / \Omega_D$. Using the expression given by Eq. (A-11), we derive the perturbed density $\tilde{n}_D$ as

$$\tilde{n}_D \simeq \frac{\pi e}{k_y T_D} \int dv_\perp^2 dv_\| f_D^M$$
$$\times \left\{ \mu_D \left[ J_0(\mu_D) J_0'(\mu_D) + \frac{\Omega_D}{\delta\omega_\alpha - k_\| v_\|} J_1(\mu_D) J_1'(\mu_D) \right] \tilde{E}_x + i \frac{\Omega_D}{\delta\omega_\alpha - k_\| v_\|} J_1^2(\mu_D) \tilde{E}_y \right\}, \tag{A-14}$$

where $f_D^M$ is the deuteron Maxwellian distribution. It may be verified that, in the limits $\mu_D \ll 1$ and $\delta\omega_\alpha \gg k_\| v_\|$, Eq. (A-14) for $\tilde{n}_D$ agrees with the result that is obtained from the fluid equations (particle and momentum conservation equations). Thus, for $b_D \equiv k_y^2 T_D / (m_D \Omega_D^2) \ll 1$ and following standard procedures,

$$\text{Im}(\tilde{n}_D) \simeq -|\tilde{n}_D| \text{Im} W_D(\lambda), \tag{A-15}$$

where $\text{Im} W_D(\lambda) = \sqrt{\pi} \lambda \exp(-\lambda^2)$ and $\lambda \equiv \delta\omega_\alpha / |k_\| V_{th}|_D$.